\documentclass[lnbip,natbib,sort&compress,numbers,square]{svmultln}

\usepackage{makeidx}  
\usepackage{listings}
\usepackage{graphicx}
\usepackage{calc}
\usepackage{color}
\usepackage{booktabs}
\usepackage{tabularx}
\usepackage{multirow}
\usepackage{amsmath}
\PassOptionsToPackage{hyphens}{url}\usepackage[pdftex, colorlinks=true, hyperfootnotes=false, hyperindex=true, plainpages=false, pagebackref=false, pdfpagelabels=true, pdfstartview=FitH, linkcolor=blue, citecolor=blue, urlcolor=blue, bookmarks=false]{hyperref}

\begin{document}
\mainmatter              
\title{Blockchain for Business Process Enactment:\\
A Taxonomy and Systematic Literature Review}
%
\titlerunning{Blockchain for Business Process Enactment -- Taxonomy and SLR}  
%
\author{Fabian Stiehle \and Ingo Weber
}
%
%
%
\institute{Software and Business Engineering, Technische Universitaet Berlin, Germany\\
\email{stiehle@campus.tu-berlin.de, ingo.weber@tu-berlin.de}}

\maketitle              
\begin{abstract}
Blockchain has been proposed to facilitate the enactment of interorganisational business processes. For such processes, blockchain can guarantee the enforcement of rules and the integrity of execution traces---without the need for a centralised trusted party.
However, the enactment of interorganisational processes pose manifold challenges.
In this work, we ask what answers the research field offers in response to those challenges.
To do so, we conduct a systematic literature review (SLR). As our guiding question, we investigate the guarantees and capabilities of blockchain-based enactment approaches.
Based on this SLR, we develop a taxonomy for blockchain-based enactment.
We find that a wide range of approaches support traceability and correctness; however, research focusing on flexibility and scalability remains nascent. For all challenges, we point towards future research opportunities.
%
\end{abstract}
\section{Introduction}
\label{sec:intro}
The enactment of a process is a central part of the business process management (BPM) lifecycle. Enactment comprises instantiation, execution, and monitoring of a process~\cite[Chapter~1.2]{weskeBusinessProcessManagement2019}. Business process management systems (BPMS), also known as workflow management systems, have long been used in \textit{intra}organisational processes to automate the enactment of business processes~\cite[Chapter~2.4]{weskeBusinessProcessManagement2019}\cite[Chapter~9.1.2]{dumasFundamentalsBusinessProcess2018}. However, in an \textit{inter}organisational setting, without central control, this is far more complex. To capture the complexity surrounding multiple autonomous distributed actors, Breu et al.\ denote such processes as \emph{living}. Such processes, they argue, make traceability, scalability, flexibility, and correctness aspects far more challenging to address~\cite{breuLivingInterorganizationalProcesses2013}. Similarly, Pourmirza et al.\ find that only 30\% of BPMS consider interorganisational aspects. For these systems, the ``autonomy of organisations'' becomes an issue. This requires trust mechanisms, dynamism, and flexibility. Furthermore, they identify standardization and interoperability issues~\cite{pourmirzaSystematicLiteratureReview2017}. 
In this setting, blockchain has been proposed to serve as a neutral ground between participants, by facilitating trust and enforcing conformance and integrity—without the introduction of a centralised trusted party~\cite{weberUntrustedBusinessProcess2016a}.
In this work, we ask what answers the research field of blockchain-based enactment offers in response to the challenges posed by interorganisational challenges. To do so, we develop a taxonomy capable of describing and classifying blockchain-based enactment approaches. We derive this taxonomy from a comprehensive systematic literature review (SLR), based on 36 selected primary studies. We find that, while blockchain is a natural fit to ensure traceability and correctness of process execution, research focusing on flexibility and scalability remains nascent. For all challenges, we point out possible future research directions. 
Following open science principles, and to enable replicability, we make the data from our SLR available---see Footnote~\ref{fn:dataset}.

\subsection{Blockchain-Based Business Process Enactment}
In an interorganisational setting, process control crosses organisational boundaries. Without central control, properties such as traceability or correctness are hard to address, e.g., how to ensure integrity and availability of event data across organisations, or how to enforce control-flow when control is distributed~\cite{breuLivingInterorganizationalProcesses2013}. With central control, the question arises which party is to host a hub or mediator component, i.e., a centralised trusted party must be introduced~\cite{weberUntrustedBusinessProcess2016a}. Blockchain technology can distribute this trust by offering ``a single \textit{logically-centralised} ledger of cryptocurrency transactions operated in an \textit{organisationally-decentralised} and \textit{physically-distributed} way''~\cite[p.~7]{weberProgrammableMoney2021}. The blockchain's ledger is in practice immutable, non-repudiable, fully transparent, and highly available~\cite[Chapter~5]{xuArchitectureBlockchainApplications2019}. Smart contracts can be used to perform arbitrary computations on the blockchain. 
As conceptualized in the first work in the field~\cite{weberUntrustedBusinessProcess2016a}, blockchain can assume control of the process, enforcing or monitoring process rules and providing an immutable process trace.

\subsection{Related Work}
Pourmirza et al.~\cite{pourmirzaSystematicLiteratureReview2017} presented a SLR of BPMS architectures; they have found that only 30\% consider interorganisational aspects. Mendling et al.~\cite{mendlingBlockchainsBusinessProcess2018a} formulated the possibilities and challenges of blockchain for BPM. Their seminal work can be seen as charting the research direction in BPM and blockchain. For enactment, they discussed the approach as outlined in Weber et al.~\cite{weberUntrustedBusinessProcess2016a}. Di Ciccio et al.~\cite{diciccioBusinessProcessMonitoring2020} discussed the possibilities of business process monitoring using blockchain, which is part of the enactment lifecycle. For blockchain and BPM as a whole, Garcia-Garcia et al.~\cite{garcia-garciaUsingBlockchainImprove2020a} conducted a SLR investigating blockchain support for the different BPM lifecycles. In contrast, we present a taxonomy and classification of enactment approaches. This allows us to provide in-depth analysis specific to enactment. To the best of the authors' knowledge, this is the first work to present a SLR and taxonomy on blockchain-based business process enactment.
\section{Methodology}
\label{sec:methodology}
A taxonomy is a classification system that produces groupings of objects based on common characteristics~\cite{nickersonMethodTaxonomyDevelopment2013}. Such a classification is integral to scientific method. In a complex field, a taxonomy can facilitate understanding and analysis. It can help navigate the research field and identify research gaps. In the field of design science, Williams et al.~\cite{williamsDesignEmergingDigital2008} note that the classification of differences provides insights into the design---and design process---of artefacts.
For taxonomy development, we follow the definitions and guidelines as outlined in Nickerson et al.~\cite{nickersonMethodTaxonomyDevelopment2013}. A taxonomy has different dimensions that can be derived inductively (i.e., empirically) or deductively (i.e., conceptually). Induction requires empirical evidence (i.e., cases to investigate), while deduction requires sound knowledge to deduce dimensions through logical reasoning. Nickerson et al. recommend the application of both methods in an iterative manner. We did so, but relied mostly on induction. To collect empirical evidence, we conducted a SLR of the field as per Kitchenham et al.~\cite{kitchenhamGuidelinesPerformingSystematic2007}, interleaved with the methods outlined by Nickerson et al.\ for taxonomy development. That is, the identified primary studies were used to inductively derive our taxonomy. Afterwards, we classified our primary studies according to the taxonomy.
\subsection{Taxonomy Development}
Following Nickerson et al.~\cite{nickersonMethodTaxonomyDevelopment2013}, at first, we have defined the users, purpose, and the meta-charateristic for our taxonomy. As \textbf{users}, we identified design science researchers. For these researchers, the \textbf{purpose} of this taxonomy is to enable the assessment of the current state of the art and future research opportunities. More specifically, which challenges of interorganisational processes have been solved by integrating blockchain, and which are still unaddressed.
A meta-characteristic is the most general and complete characteristic from which all dimensions are derived~\cite{nickersonMethodTaxonomyDevelopment2013}. This characteristic can be thought of as the starting point for taxonomy development. Our \textbf{meta-characteristic} is comprised of the \textit{guarantees and capabilities of blockchain-based process enactment}. Distributing trust is the central reason for introducing blockchain technology to process enactment. Blockchain establishes trust by providing certain guarantees, such as the immutability of the ledger. Therefore, the offered guarantees were of central interest to our research. In addition, we investigated the capabilities of approaches, such as resource allocation and process flexibility. The meta-characteristic also served as the guiding research question for our SLR.

\subsection{Systematic Literature Review}
Through early exploratory searches, we could not deduce a concise common terminology for blockchain-based business process enactment. Thus, we decided to conduct a search with a set of broad search keywords, connecting terms of business process management with blockchain, and then apply more restrictive exclusion criteria. To limit the search results (given that blockchain constitutes a buzzword mentioned in many works), we restricted the search to the title of studies. The search string is presented in Listing~\ref{ls:search}.

\begin{lstlisting}[breaklines,basicstyle=\footnotesize\ttfamily, caption={Search string. Note, in order to save space, here we implicitly mean both singular and plural versions of each search keyword.},label=ls:search, language=Pascal,escapeinside={(*@}{@*)}]
("blockchain" OR "smart contract" OR "DLT" OR 
"Distributed Ledger Technology") (*@ \textbf{AND} @*)
("bpm" OR "business process" OR "choreography" OR 
"workflow"))
\end{lstlisting}
To account for the fast research pace in which blockchain is evolving, we also considered pre-prints and conference papers. There is evidence that \emph{Google Scholar}\footnote{\url{https://scholar.google.com}, accessed 2022-05-30.} performs especially well in such scenarios~\cite{martin-martinGoogleScholarWeb2018}, which made it our tool of choice. The initial search was conducted on the 2022-03-10 and yielded 186 entries. A full list of applied inclusion and exclusion criteria is given in Table~\ref{tab:incex} below. In the first pass, we excluded works based on publication type and title; in the second we examined the abstract. Finally, we conducted a full reading.
After applying our exclusion criteria, we obtained 30 studies. We then performed backward snowballing. To limit the scope of the study, we did not conduct a full forward snowballing. Due to our broad search keywords, we expected forward snowballing to only yield a large set of irrelevant or already reviewed studies. We confirmed this expectation for our two most cited primary studies, and indeed found no relevant additional studies. Trough backward snowballing, we obtained an additional six studies, leading to a final primary study set of 36 studies. The full process, each pass, and the application of the exclusion criteria is made transparent in our published data set.\footnote{Replication package available at: \url{https://github.com/fstiehle/SLR-blockchain-BP-execution}; for convenience, we also include a hosted interactive spreadsheet of our SLR at \url{https://tubcloud.tu-berlin.de/s/M8JQtaRX5JkjXXZ}.
\label{fn:dataset}}

\begin{table}[b]
\centering
\begin{tabularx}{\linewidth}{lX}
\toprule
\textit{Inclusion} & The study presents an approach in the field of blockchain-based business processes enactment. \\
\midrule 
\textit{Exclusion} & 1. The study presents a domain specific application (not meant for general business processes). \\
& 2. The study is a theoretical work, or a non-technical work, it does not present and evaluate a research artefact such as a execution or monitoring engine. \\
& 3. The study is a tertiary study, i.e., it is a review or overview of other contributions. \\
& 4. The study is illegible, i.e, not written in English or containing heavy spelling mistakes. \\
\bottomrule
\end{tabularx}
\caption{Inclusion and exclusion criteria.}
\label{tab:incex}
\end{table}


\section{A Taxonomy of Blockchain-Based Enactment}
\subsection{Overview}
\begin{figure}[h]
    \centering
    \includegraphics[width=\linewidth]{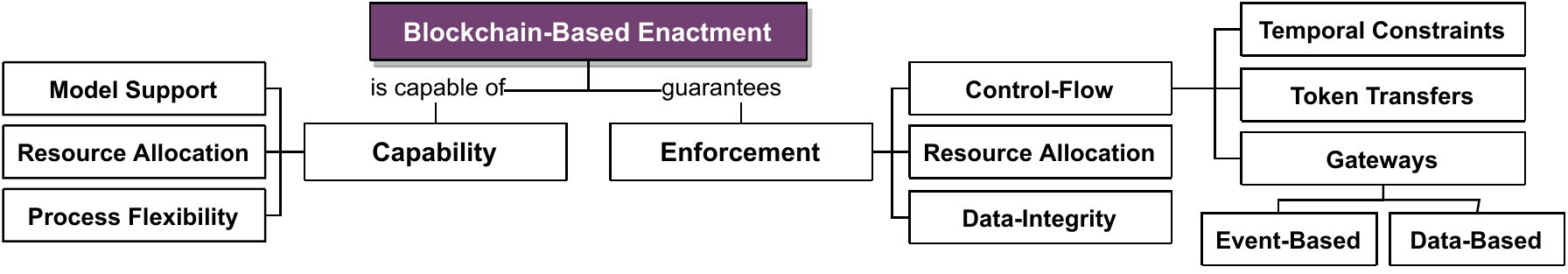}
    \caption{Overview of our taxonomy of blockchain-based process enactment.}
    \label{fig:taxonomy}
\end{figure}
\noindent By investigating our primary studies and following the methodology as per Section~\ref{sec:methodology}, we arrived at the taxonomy depicted in Figure~\ref{fig:taxonomy}. When assessing blockchain-based applications, it is important to differentiate between application-specific properties and properties inherited from the employed blockchain. In our taxonomy, dimensions are kept independent from the chosen blockchain; this allows an independent assessment. The decision which blockchain platform to use is a different, but interrelated design decision~\cite[Chapter~6.3]{xuArchitectureBlockchainApplications2019}. We have structured our taxonomy into supported capabilities and enforced guarantees. To improve readability, we defer the detailed introduction of our dimensions to the presentation of our classification results in Section~\ref{sec:results}. For capabilities, we capture with model support, which notation was chosen to represent the business process; with resource allocation capability, we differentiate between resource allocation strategies; with process flexibility capability, we capture how studies approach flexibility.
For enforcement, we find that control-flow, resource allocation, and data-integrity aspects are enforced on-chain. 
Given the prevalence of control-flow, we subdivide this aspect further (See Section~\ref{sec:results} below).
%
%
\subsection{Dimensions and Classification Results}
\label{sec:results}
We detail the dimensions of our taxonomy and present the results of the classification our our primary studies.
%
\subsubsection{Capabilities.}
First, we explore the supported capabilities, which are summarised in Table~\ref{tab:capabilities}.
\paragraph{Model Support.}
The vast majority (70\%) of studies are BPMN based. Notably, there is no close second. 
In terms of BPMN, 13 studies support the process, eight the choreography, and four the collaboration diagram.
\begin{table}[t]
\scriptsize
\centering
\begin{tabular}{lrl}
\toprule
\textbf{Capability} & Number (\% of total) & Reference list \\ 
\midrule
\textbf{Model Support} & \textbf{36 (100\%)} & \\
\midrule
BPMN process diagram & 13 (36\%) & \cite{garcia-banuelosOptimizedExecutionBusiness2017a,lopez-pintadoCaterpillarBusinessProcess2019a,lopez-pintadoDynamicRoleBinding2019a,luIntegratedModelDriven2021a,nakamuraInterorganizationalBusinessProcesses2018a,lopez-pintadoInterpretedExecutionBusiness2019a,lopez-pintadoControlledFlexibilityBlockchainbased2022a,mercenneBlockchainStudioRolebased2018a,alvesExploringBlockchainTechnology2020a,brahemRunningTransactionalBusiness2020a,sturmLeanArchitectureBlockchain2018,falaziProcessbasedCompositionPermissioned2019,abidModellingExecutingTimeAware2020} \\
BPMN choreography diagram & 8 (22\%) & \cite{ladleifModelingEnforcingBlockchainBased2019,corradiniEngineeringTrustableChoreographybased2020a,weberUntrustedBusinessProcess2016a,prybilaRuntimeVerificationBusiness2020a,corradiniEngineeringTrustableAuditable2022a,lichtensteinDataDrivenProcessChoreography2020,loukilDecentralizedCollaborativeBusiness2021a,corradiniModeldrivenEngineeringMultiparty2021a} \\
BPMN collaboration diagram & 4 (11\%) & \cite{klingerUpgradeabilityConceptCollaborative2020a,morales-sandovalBlockchainSupportExecution2021a,klingerBlockchainbasedCrossOrganizationalExecution2020,sturmBlockchainbasedResourceawareProcess2019} \\
Undescribed model & 4 (11\%) & \cite{boreUsingBlockchainBased2019a,naganoRELIABLEARCHITECTURECROSS2020a,naganoBLOCKCHAINBASEDCROSSa,osterlandImplementationBusinessProcess2020} \\
YAWL & 2 (5\%) & \cite{adamsFlexibleIntegrationBlockchain2020a,evermannAdaptingWorkflowManagement2020a} \\
Petri net & 2 (5\%) & \cite{evermannWorkflowManagementProofofWork2021a,evermannWorkflowManagementBFT2020a} \\
Other (DCR, DEMO, and GSM) & 3 (8\%) & \cite{madsenCollaborationAdversariesDistributed2018a,silvaDecentralizedEnforcementBusiness2019,meroniTrustedArtifactDrivenProcess2019} \\
\midrule
\textbf{Resource Allocation} & \textbf{21 (58\%)} \\
\midrule
Role-based & 13 (36\%) & \cite{ladleifModelingEnforcingBlockchainBased2019,corradiniEngineeringTrustableChoreographybased2020a,weberUntrustedBusinessProcess2016a,boreUsingBlockchainBased2019a,corradiniEngineeringTrustableAuditable2022a,mercenneBlockchainStudioRolebased2018a,silvaDecentralizedEnforcementBusiness2019,klingerUpgradeabilityConceptCollaborative2020a,morales-sandovalBlockchainSupportExecution2021a,loukilDecentralizedCollaborativeBusiness2021a,corradiniModeldrivenEngineeringMultiparty2021a,klingerBlockchainbasedCrossOrganizationalExecution2020,sturmBlockchainbasedResourceawareProcess2019} \\
Direct & 5 (14\%) & \cite{madsenCollaborationAdversariesDistributed2018a,prybilaRuntimeVerificationBusiness2020a,meroniTrustedArtifactDrivenProcess2019,sturmLeanArchitectureBlockchain2018,osterlandImplementationBusinessProcess2020} \\
Dynamic & 3 (8\%)& \cite{lopez-pintadoDynamicRoleBinding2019a,lopez-pintadoInterpretedExecutionBusiness2019a,lopez-pintadoControlledFlexibilityBlockchainbased2022a} \\
\midrule
\textbf{Process Flexibility} & \textbf{5 (13\%)} & \\
\midrule
Looseness & 3 (8\%) & \cite{madsenCollaborationAdversariesDistributed2018a,lopez-pintadoControlledFlexibilityBlockchainbased2022a,meroniTrustedArtifactDrivenProcess2019} \\
Adaptation & 1 (2\%) & \cite{lopez-pintadoControlledFlexibilityBlockchainbased2022a}\\
Evolution & 1 (2\%) & \cite{klingerUpgradeabilityConceptCollaborative2020a} \\
\midrule
\bottomrule
\end{tabular}
\caption{Classification of capabilities.}
\label{tab:capabilities}
\end{table}
\paragraph{Resource Allocation.}
\label{sec:resourcealloc}
Resource allocation assigns a process resource to a task~\cite[Chapter~10.5]{dumasFundamentalsBusinessProcess2018}. In blockchain-based enactment, a resource is typically identified by a blockchain account address. As blockchain transactions must be signed, a resource’s involvement in a task cannot be repudiated (assuming the secrecy of their private key).
21 studies support resource allocation in general. Of these, we can differentiate between direct (five studies) and role-based (13 studies). A direct allocation binds a blockchain address directly to a task. Role-based allocation allows some indirection by assigning addresses to roles and roles to tasks. Only three allow a more dynamic strategy, which was first presented in~\cite{lopez-pintadoDynamicRoleBinding2019a}. These dynamic variants allow to specify the conditions for resource allocation in a so-called binding policy. For a given role, a participant can be nominated. Constraints can require that a participant must (or must not) already be bound to certain other roles. Endorsement constraints specify when and which other participants can vote on a nomination.
\paragraph{Process Flexibility.}
Process flexibility is essential for supporting less predictable processes. Reichert et al.\ characterised four flexibility needs: variability, looseness, adaptation, and evolution~\cite[Chapter~3]{reichertEnablingFlexibilityProcessAware2012}. We find that only five studies support flexibility needs. Looseness is supported by three studies. Two \cite{madsenCollaborationAdversariesDistributed2018a, meroniTrustedArtifactDrivenProcess2019} support looseness by using declarative models; these are  loosely specified, providing more flexibility by only modelling constraints~\cite[Chapter~12]{reichertEnablingFlexibilityProcessAware2012}. 
L{\'o}pez-Pintado et al.~\cite{lopez-pintadoControlledFlexibilityBlockchainbased2022a} support looseness and adaptation. Their approach allows late modelling: subprocesses can be modelled during run time. Furthermore, certain process elements can be adapted during run time. This is accompanied by an agreement policy, which allows to specify the participants that are allowed to adapt process elements and the conditions that must be met. For the adapted process, they can guarantee deadlock freeness.
Klinger et al.~\cite{klingerUpgradeabilityConceptCollaborative2020a} support process evolution. They implement blockchain design patterns (registry and proxy patterns) that enable the versioning of processes. These patterns decouple logic from data and allow logic to be updated. The approach includes a voting mechanism, which allows participants to vote on new process versions.
%
%
%
\subsubsection{Enforcement Guarantees.}
Different perspectives of a process can be enforced on the blockchain. We find that control-flow, data-integrity, and resource allocation are enforced on-chain. Table \ref{tab:enforcement} gives an overview of our classification result. We can observe a clear focus on control-flow enforcement (31 studies) over monitoring (5 studies).\footnote{We define monitoring as approaches where the control-flow is not enforced, but the process trace is still committed to the blockchain to ensure the integrity of the trace.} Control-flow enforcement can be \textit{data-based} (15 studies) or \textit{event-based} (16 studies). Data-based enforcement enables, based on instance data, the evaluation of gateway conditions to automatically allow or disallow certain branches in the model. Event-based enforcement, on the other hand, can only enforce the semantics of the gateway (e.g., branching semantics of AND or XOR gateways), but not evaluate dynamic conditions.

Additionally, public blockchains enable the enforcement of token transfers (six studies). That is, certain behaviour may prompt automatic transfer of crypto tokens. Beyond fungible tokens (five studies), only Lu et al.~\cite{luIntegratedModelDriven2021a} support the modelling and transfer of non-fungible tokens, which are integral for asset management.

Finally, only Ladleif et al.~\cite{ladleifModelingEnforcingBlockchainBased2019} and Abid et al.~\cite{abidModellingExecutingTimeAware2020} allow to enforce temporal constraints. These constraints are based on the block timestamp.
A blockchain network has no strong notion of a synchronised clock, the close world assumption and the transaction-driven nature of blockchain do not allow to access external time information or to continuously monitor an internal clock~\cite{ladleifTimeBlockchainBasedProcess2020}. The block timestamp is the only readily available traditional notion of time on the chain; it is, however, of limited accuracy and can---to a certain extend---be manipulated by the block creator~\cite{ladleifTimeBlockchainBasedProcess2020}.

We listed 21 studies that support the allocation of resources to tasks. Most studies (19 studies) enforce this allocation by implementing authorisation mechanisms: only the allocated blockchain address can perform the task. In contrast, Prybila et al.~\cite{prybilaRuntimeVerificationBusiness2020a} and Meroni et al.~\cite{meroniTrustedArtifactDrivenProcess2019} present monitoring approaches that only guarantee the authenticity of the resource that has performed the task, they do not enforce authorisation. Lastly, all approaches make use of the integrity guarantee of blockchain to store the process trace. 27 studies allow the storage of instance data and five store a serialised version of the (original) process model on the blockchain.
\begin{table}[t]
\scriptsize
\centering
\begin{tabular}{lrl}
\toprule
\textbf{Enforcement Guarantee} ~ & Number (\% of total) & Reference list \\ 
\midrule
\textbf{Control-Flow} & \textbf{31 (86\%)} & \\
\midrule
Event-based gateways & 16 (44\%) & \cite{weberUntrustedBusinessProcess2016a,madsenCollaborationAdversariesDistributed2018a,boreUsingBlockchainBased2019a,nakamuraInterorganizationalBusinessProcesses2018a,silvaDecentralizedEnforcementBusiness2019,klingerUpgradeabilityConceptCollaborative2020a,evermannWorkflowManagementProofofWork2021a,loukilDecentralizedCollaborativeBusiness2021a,evermannAdaptingWorkflowManagement2020a,evermannWorkflowManagementBFT2020a,naganoRELIABLEARCHITECTURECROSS2020a,naganoBLOCKCHAINBASEDCROSSa,klingerBlockchainbasedCrossOrganizationalExecution2020,sturmLeanArchitectureBlockchain2018,sturmBlockchainbasedResourceawareProcess2019,osterlandImplementationBusinessProcess2020} \\
Data-based gateways & 15 (42\%) & \cite{ladleifModelingEnforcingBlockchainBased2019,garcia-banuelosOptimizedExecutionBusiness2017a,corradiniEngineeringTrustableChoreographybased2020a,lopez-pintadoCaterpillarBusinessProcess2019a,lopez-pintadoDynamicRoleBinding2019a,luIntegratedModelDriven2021a,lopez-pintadoInterpretedExecutionBusiness2019a,corradiniEngineeringTrustableAuditable2022a,lopez-pintadoControlledFlexibilityBlockchainbased2022a,mercenneBlockchainStudioRolebased2018a,lichtensteinDataDrivenProcessChoreography2020,morales-sandovalBlockchainSupportExecution2021a,corradiniModeldrivenEngineeringMultiparty2021a,brahemRunningTransactionalBusiness2020a,abidModellingExecutingTimeAware2020} \\
Token transfers & 6 (17\%) & \cite{corradiniEngineeringTrustableChoreographybased2020a,weberUntrustedBusinessProcess2016a,luIntegratedModelDriven2021a,corradiniEngineeringTrustableAuditable2022a,corradiniModeldrivenEngineeringMultiparty2021a,falaziProcessbasedCompositionPermissioned2019} \\
Temporal constraints & 2 (6\%) & \cite{ladleifModelingEnforcingBlockchainBased2019,abidModellingExecutingTimeAware2020} \\
\midrule
\textbf{Resource Allocation} & \textbf{19 (53\%)} & \cite{ladleifModelingEnforcingBlockchainBased2019,corradiniEngineeringTrustableChoreographybased2020a,weberUntrustedBusinessProcess2016a,madsenCollaborationAdversariesDistributed2018a,boreUsingBlockchainBased2019a,lopez-pintadoDynamicRoleBinding2019a,lopez-pintadoInterpretedExecutionBusiness2019a,corradiniEngineeringTrustableAuditable2022a,lopez-pintadoControlledFlexibilityBlockchainbased2022a,mercenneBlockchainStudioRolebased2018a,silvaDecentralizedEnforcementBusiness2019,klingerUpgradeabilityConceptCollaborative2020a,morales-sandovalBlockchainSupportExecution2021a,loukilDecentralizedCollaborativeBusiness2021a,corradiniModeldrivenEngineeringMultiparty2021a,klingerBlockchainbasedCrossOrganizationalExecution2020,sturmLeanArchitectureBlockchain2018,sturmBlockchainbasedResourceawareProcess2019,osterlandImplementationBusinessProcess2020} \\
\midrule
\textbf{Data-Integrity} & \textbf{36 (100\%)} & \\
\midrule
Execution trace & 36 (100\%) & \cite{ladleifModelingEnforcingBlockchainBased2019,garcia-banuelosOptimizedExecutionBusiness2017a,corradiniEngineeringTrustableChoreographybased2020a,weberUntrustedBusinessProcess2016a,madsenCollaborationAdversariesDistributed2018a,boreUsingBlockchainBased2019a,lopez-pintadoCaterpillarBusinessProcess2019a,lopez-pintadoDynamicRoleBinding2019a,prybilaRuntimeVerificationBusiness2020a,luIntegratedModelDriven2021a,nakamuraInterorganizationalBusinessProcesses2018a,lopez-pintadoInterpretedExecutionBusiness2019a,corradiniEngineeringTrustableAuditable2022a,lopez-pintadoControlledFlexibilityBlockchainbased2022a,mercenneBlockchainStudioRolebased2018a,silvaDecentralizedEnforcementBusiness2019,adamsFlexibleIntegrationBlockchain2020a,meroniTrustedArtifactDrivenProcess2019,alvesExploringBlockchainTechnology2020a,klingerUpgradeabilityConceptCollaborative2020a,lichtensteinDataDrivenProcessChoreography2020,morales-sandovalBlockchainSupportExecution2021a,evermannWorkflowManagementProofofWork2021a,loukilDecentralizedCollaborativeBusiness2021a,corradiniModeldrivenEngineeringMultiparty2021a,evermannAdaptingWorkflowManagement2020a,evermannWorkflowManagementBFT2020a,naganoRELIABLEARCHITECTURECROSS2020a,brahemRunningTransactionalBusiness2020a,naganoBLOCKCHAINBASEDCROSSa,klingerBlockchainbasedCrossOrganizationalExecution2020,sturmLeanArchitectureBlockchain2018,falaziProcessbasedCompositionPermissioned2019,sturmBlockchainbasedResourceawareProcess2019,abidModellingExecutingTimeAware2020,osterlandImplementationBusinessProcess2020} \\
Instance data & 27 (75\%) & \cite{ladleifModelingEnforcingBlockchainBased2019,garcia-banuelosOptimizedExecutionBusiness2017a,corradiniEngineeringTrustableChoreographybased2020a,weberUntrustedBusinessProcess2016a,boreUsingBlockchainBased2019a,lopez-pintadoCaterpillarBusinessProcess2019a,lopez-pintadoDynamicRoleBinding2019a,prybilaRuntimeVerificationBusiness2020a,luIntegratedModelDriven2021a,lopez-pintadoInterpretedExecutionBusiness2019a,corradiniEngineeringTrustableAuditable2022a,lopez-pintadoControlledFlexibilityBlockchainbased2022a,mercenneBlockchainStudioRolebased2018a,silvaDecentralizedEnforcementBusiness2019,adamsFlexibleIntegrationBlockchain2020a,meroniTrustedArtifactDrivenProcess2019,alvesExploringBlockchainTechnology2020a,lichtensteinDataDrivenProcessChoreography2020,evermannWorkflowManagementProofofWork2021a,corradiniModeldrivenEngineeringMultiparty2021a,evermannAdaptingWorkflowManagement2020a,evermannWorkflowManagementBFT2020a,naganoRELIABLEARCHITECTURECROSS2020a,brahemRunningTransactionalBusiness2020a,naganoBLOCKCHAINBASEDCROSSa,abidModellingExecutingTimeAware2020,osterlandImplementationBusinessProcess2020} \\
Process model & 5 (14\%) & \cite{boreUsingBlockchainBased2019a,meroniTrustedArtifactDrivenProcess2019,evermannWorkflowManagementProofofWork2021a,evermannAdaptingWorkflowManagement2020a,evermannWorkflowManagementBFT2020a} \\
\midrule
\bottomrule
\end{tabular}
\caption{Classification of enforcement guarantees.}
\label{tab:enforcement}
\end{table}
\subsubsection{Methods.}
\label{sec:methods}
Beyond our taxonomy, we investigate the evaluation methods employed. A summary is presented in  Table~\ref{tab:methods}. 
Most works evaluate their approach using Ethereum (24 studies). Nine studies use Hyperledger Fabric. Four present a custom blockchain implementation and only two studies consider Bitcoin. Finally, Corradini et al.~\cite{corradiniModeldrivenEngineeringMultiparty2021a} and Falazi et al.~\cite{falaziProcessbasedCompositionPermissioned2019} present artefacts for multiple blockchains.
\begin{table}[t]
\scriptsize
\centering
\begin{tabularx}{\textwidth}{lrX}
\toprule
\textbf{Method} & Number (\% of total) & Reference list \\ 
\midrule
\textbf{Blockchain Selection} & \textbf{36 (100\%)} & \\
\midrule
Ethereum & 24 (67\%) & \cite{ladleifModelingEnforcingBlockchainBased2019,garcia-banuelosOptimizedExecutionBusiness2017a,corradiniEngineeringTrustableChoreographybased2020a,weberUntrustedBusinessProcess2016a,madsenCollaborationAdversariesDistributed2018a,lopez-pintadoCaterpillarBusinessProcess2019a,lopez-pintadoDynamicRoleBinding2019a,luIntegratedModelDriven2021a,lopez-pintadoInterpretedExecutionBusiness2019a,corradiniEngineeringTrustableAuditable2022a,lopez-pintadoControlledFlexibilityBlockchainbased2022a,mercenneBlockchainStudioRolebased2018a,meroniTrustedArtifactDrivenProcess2019,klingerUpgradeabilityConceptCollaborative2020a,lichtensteinDataDrivenProcessChoreography2020,morales-sandovalBlockchainSupportExecution2021a,loukilDecentralizedCollaborativeBusiness2021a,corradiniModeldrivenEngineeringMultiparty2021a,brahemRunningTransactionalBusiness2020a,klingerBlockchainbasedCrossOrganizationalExecution2020,sturmLeanArchitectureBlockchain2018,falaziProcessbasedCompositionPermissioned2019,sturmBlockchainbasedResourceawareProcess2019,abidModellingExecutingTimeAware2020} \\ 
Hyperledger Fabric & 9 (25\%) & \cite{boreUsingBlockchainBased2019a,nakamuraInterorganizationalBusinessProcesses2018a,silvaDecentralizedEnforcementBusiness2019,adamsFlexibleIntegrationBlockchain2020a,alvesExploringBlockchainTechnology2020a,corradiniModeldrivenEngineeringMultiparty2021a,naganoRELIABLEARCHITECTURECROSS2020a,naganoBLOCKCHAINBASEDCROSSa,falaziProcessbasedCompositionPermissioned2019} \\
Custom implementation & 4 (11\%) & \cite{evermannWorkflowManagementProofofWork2021a,evermannAdaptingWorkflowManagement2020a,evermannWorkflowManagementBFT2020a,osterlandImplementationBusinessProcess2020} \\
Bitcoin & 2 (6\%) & \cite{prybilaRuntimeVerificationBusiness2020a,falaziProcessbasedCompositionPermissioned2019} \\
\midrule
\textbf{Evaluation Criteria} & \textbf{33 (92\%)} & \\
\midrule
Cost & 23 (64\%) & \cite{ladleifModelingEnforcingBlockchainBased2019,garcia-banuelosOptimizedExecutionBusiness2017a,corradiniEngineeringTrustableChoreographybased2020a,weberUntrustedBusinessProcess2016a,madsenCollaborationAdversariesDistributed2018a,lopez-pintadoCaterpillarBusinessProcess2019a,lopez-pintadoDynamicRoleBinding2019a,prybilaRuntimeVerificationBusiness2020a,luIntegratedModelDriven2021a,nakamuraInterorganizationalBusinessProcesses2018a,lopez-pintadoInterpretedExecutionBusiness2019a,corradiniEngineeringTrustableAuditable2022a,lopez-pintadoControlledFlexibilityBlockchainbased2022a,mercenneBlockchainStudioRolebased2018a,meroniTrustedArtifactDrivenProcess2019,klingerUpgradeabilityConceptCollaborative2020a,lichtensteinDataDrivenProcessChoreography2020,evermannWorkflowManagementProofofWork2021a,loukilDecentralizedCollaborativeBusiness2021a,corradiniModeldrivenEngineeringMultiparty2021a,klingerBlockchainbasedCrossOrganizationalExecution2020,sturmLeanArchitectureBlockchain2018,sturmBlockchainbasedResourceawareProcess2019} \\
Qualitative discussion & 18 (50\%) & \cite{ladleifModelingEnforcingBlockchainBased2019,weberUntrustedBusinessProcess2016a,prybilaRuntimeVerificationBusiness2020a,corradiniEngineeringTrustableAuditable2022a,silvaDecentralizedEnforcementBusiness2019,adamsFlexibleIntegrationBlockchain2020a,meroniTrustedArtifactDrivenProcess2019,alvesExploringBlockchainTechnology2020a,evermannWorkflowManagementProofofWork2021a,loukilDecentralizedCollaborativeBusiness2021a,corradiniModeldrivenEngineeringMultiparty2021a,evermannAdaptingWorkflowManagement2020a,evermannWorkflowManagementBFT2020a,naganoRELIABLEARCHITECTURECROSS2020a,naganoBLOCKCHAINBASEDCROSSa,sturmLeanArchitectureBlockchain2018,falaziProcessbasedCompositionPermissioned2019,sturmBlockchainbasedResourceawareProcess2019} \\
Correctness & 9 (25\%) & \cite{garcia-banuelosOptimizedExecutionBusiness2017a,weberUntrustedBusinessProcess2016a,lopez-pintadoCaterpillarBusinessProcess2019a,prybilaRuntimeVerificationBusiness2020a,luIntegratedModelDriven2021a,meroniTrustedArtifactDrivenProcess2019,morales-sandovalBlockchainSupportExecution2021a,loukilDecentralizedCollaborativeBusiness2021a,osterlandImplementationBusinessProcess2020} \\
Throughput & 3 (8\%) & \cite{garcia-banuelosOptimizedExecutionBusiness2017a,evermannWorkflowManagementBFT2020a,osterlandImplementationBusinessProcess2020} \\
Finality & 3 (8\%) & \cite{weberUntrustedBusinessProcess2016a,prybilaRuntimeVerificationBusiness2020a,corradiniModeldrivenEngineeringMultiparty2021a} \\
\midrule
\bottomrule
\end{tabularx}
\caption{Employed methods.}
\label{tab:methods}
\end{table}
\begin{figure}[t]
    \centering
    \includegraphics[width=.7\linewidth]{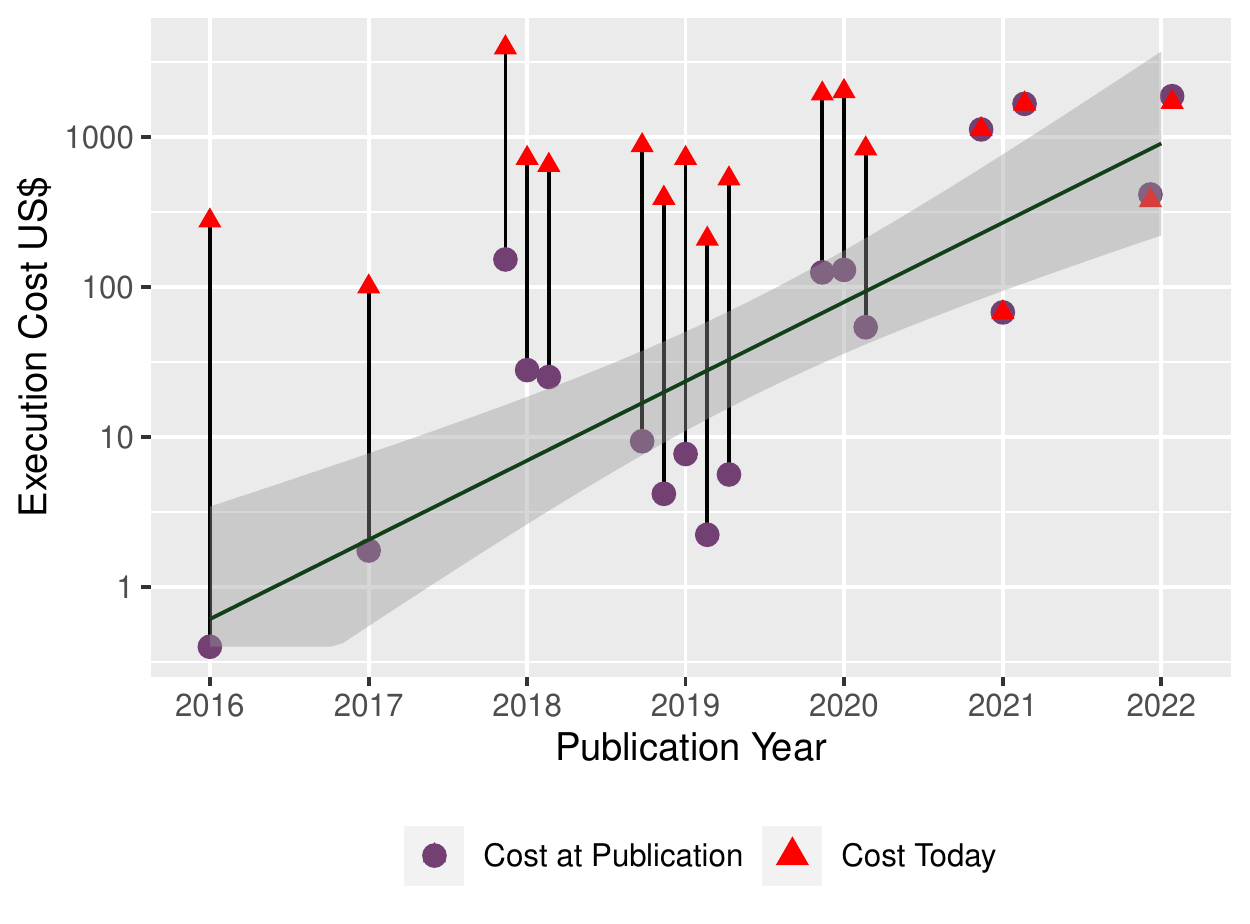}
    \caption{Execution cost on Ethereum at publication and today.}
    \label{fig:cost}
\end{figure}
In terms of evaluation, cost is the most frequently regarded metric (23 studies). For ten studies, cost is the sole focus of the evaluation.

\noindent Indeed, this concern is understandable when considering the cost of public blockchain compared to more traditional computing~\cite{rimbaQuantifyingCostDistrust2020a}. Most (19 studies out of 23) report cost based on Ethereum gas\footnote{See \textit{Gas and fees}, \url{https://ethereum.org/en/developers/docs/gas/}, accessed 2022-05-30.}. 
Transaction fees on Ethereum have increased significantly along with the popularity of the network. While Weber et al.~\cite{weberUntrustedBusinessProcess2016a} were still able to conduct evaluation experiments on the main network of Ethereum, more recent works resort to test networks or private deployments. We show this development in Figure~\ref{fig:cost}, where we compare publication year and execution cost for a repeated instance execution, excluding the cost that only incur once (e.g., deployment or configuration). As a point, we depict the original cost in US\$ at the time of publication. This price is based on the average gas cost and exchange rate of the publication year.\footnote{Our script and data sets used for calculation can be found at: \url{https://github.com/fstiehle/SLR-blockchain-BP-execution}. Historical data was taken from Etherscan (\url{https://etherscan.io/chart/gasprice}) and Yahoo Finance (\url{https://finance.yahoo.com/quote/ETH-USD/}, both accessed 2022-06-01.} The point is connected vertically with a triangle, which represents the cost the same execution would incur taken the average gas cost and exchange rate of 2021. We can observe that the cost has risen significantly over time. The result is a projected mean cost of \$1010 for a singular instance run in 2021. No approach would lie significantly below \$100 for one run.
As a result, most recent works argue for the use of a private network in almost all scenarios.

Besides cost, 18 studies discuss qualitative aspects (e.g., privacy or trust) and nine studies evaluate the correctness of their approach. For correctness, they investigate whether non-conforming traces are prevented (enforcing approaches) or detected (monitoring approaches) correctly.\footnote{We here need to remark that~\cite{sturmLeanArchitectureBlockchain2018} is not handling the claimed subset of BPMN correctly; as noted in~\cite[Sect. 2.2]{lopez-pintadoCaterpillarBusinessProcess2019a}, the OR join is handled incorrectly.}
Notably, finality\footnote{We consider finality as the time it takes a transaction to be durably committed with a certain probability $e$.} and throughput are only investigated by three studies each. 
Lastly, 18 studies have published their code and eleven have made a replication package available. 
\section{Discussion: Challenges and Future Research Directions}
%
We discuss the guarantees and capabilities of blockchain-based enactment in the light of the challenges of interorganisational processes (see Section~\ref{sec:intro}): interoperability~\cite{pourmirzaSystematicLiteratureReview2017}, traceability, scalability, flexibility, and correctness~\cite{breuLivingInterorganizationalProcesses2013}.
\subsection{Interoperability}
Blockchain can facilitate interoperability, as participants share the same execution environment. Based on our review, we can observe two opportunities for future research: supporting the data perspective, and cross-chain compatibility. 
%
%
%
\subsubsection{Data perspective.}
A shared understanding of data is essential for participants; for example, to connect the local data model or assess security and privacy implications~\cite{meyerAutomatingDataExchange2015}. Thus, the data perspective needs to be suitably modelled. All approaches, with the notable exception of Lu et al.~\cite{luIntegratedModelDriven2021a}, which use a UML class diagram to model non-fungible-tokens, require blockchain-specific code snippets (e.g., solidity code) to express data types or data-based gateway conditions. In the future, we envision the integration of more comprehensive and platform-independent data models.

\subsubsection{Cross-chain compatibility.}

Only \cite{corradiniModeldrivenEngineeringMultiparty2021a,falaziProcessbasedCompositionPermissioned2019} present artefacts for multiple blockchains. However, the choice of the blockchain platform is application-specific~\cite[Chapter~6.3]{xuArchitectureBlockchainApplications2019}. 
In terms of cross-chain compatibility, we envision three lines of research. First, when and how are multi-chain deployments a suitable implementation choice, and which basic requirements arise (e.g., cross-chain guarantees on integrity)?
Second, the creation of artefacts for different blockchain platforms from one process model. 
Third, as discussed in~\cite{ladleifArchitectureMultichainBusiness2020}, the execution of parts of the same process instance on different blockchain platforms. 
%
\subsection{Traceability \& Correctness}
Blockchain is a natural fit to ensure traceability and correctness of execution.
Event data stored on the blockchain ledger is immutable and globally traceable. The integrity of this data can be enforced without introducing a centralised trusted party.
To ensure correctness, the model-driven engineering paradigm is typically applied. 
This allows to generate well-tested artefacts following best practices. Consequently, all approaches either enforce the control-flow on-chain or allow the monitoring of the control-flow by committing the process trace to the ledger. In the following, we outline two opportunities for future research: dispute resolution and the extension of enforcement guarantees.
\subsubsection{Dispute resolution.}
Based on the immutable process trace, blockchain is envisioned to enable the resolution of contractual disputes between process participants (see e.g.,~\cite{weberUntrustedBusinessProcess2016a}). However, no approach details a dispute resolution process, nor is its facilitation supported. It is unclear in which state the process remains once a dispute is raised. 
We expect that research conducted in this direction could provide real benefit to organisations.
However, integrating the resolution of disputes may prove to be challenging and require different escalation levels~\cite{miglioriniRiseEnforceableBusiness2019a}. Furthermore, it remains unclear whether blockchain traces would be accepted in a litigation process. A dispute resolution process could also include incentive mechanisms, facilitating the honest behaviour of participants and penalising malicious behaviour. Such research, would have to be conducted in an interdisciplinary context, including law and economic disciplines.
\subsubsection{Enforcement guarantees.}
While control-flow, resource allocation, and integrity aspects are supported, we expect that organisations would benefit from the blockchain-based enforcement of other process related rules. Our taxonomy remains extensible to further enforcement dimensions. For example, resource allocation is a complex decision problem into which many characteristics can be factored in~\cite[Chapter~10.5]{dumasFundamentalsBusinessProcess2018}. 
Enforcing these rules on-chain would make the allocation process transparent and globally enforceable. Currently, most works focus on role-based allocation, but it remains intransparent why a certain participant was allocated to a role. 

Lastly, while most primary studies support on-chain enforcement, only nine evaluate the correctness of this enforcement capability. A more stringent evaluation, or even formal correctness proofs of the enforcement capability should be a central concern for the field, as this is the basis for all guarantees offered.

\subsection{Flexibility \& Scalability}
While traceability and correctness aspects are already well supported, enabling flexibility and scalability remains a challenge. We see three major research opportunities: controlled flexibility, comprehensive performance studies, and enactment on public blockchains.

\subsubsection{Controlled flexibility.}
Only five studies address flexibility challenges. Introducing flexibility in blockchain applications is a challenge due to the ledger's immutability. Furthermore, introducing flexibility capabilities may lead to trust concerns and correctness issues~\cite{lopez-pintadoControlledFlexibilityBlockchainbased2022a}. Participants must be convinced that flexibility will not introduce uncertainties beyond their control, otherwise it will undermine traceability and correctness guarantees. Recharting the development of traditional enactment approaches, we can observe that the main focus, so far, has been on predictable processes. Addressing the challenges surrounding unpredictable processes and integrating different techniques to support variability, looseness, adaptation, and evolution~\cite{reichertEnablingFlexibilityProcessAware2012}---but in a controlled manner---remains a line for future research.

\subsubsection{Comprehensive performance studies.}
Across all studies, the most prominent evaluation goal is to demonstrate (low) gas cost. Gas cost can give an indication on throughput scalability on a public or private blockchain. The notion of gas has been introduced in Ethereum to calculate transaction fees. The goal was to control network propagation and storage requirements.\footnote{Blockchain and Mining, Ethereum Whitepaper, \url{https://ethereum.org/en/whitepaper/##blockchain-and-mining}, accessed 2022-06-01.} However, other factors also play a crucial role. In our set of primary studies, scalability factors beyond gas cost are rarely explored. While a lot of performance properties depend upon the underlying blockchain platform, many use cases, especially private deployments, would benefit from more comprehensive performance studies. For example, in a private blockchain, with a few participants, gas cost may be not of paramount importance. The choice and configuration of a blockchain network is a complex trade-off between different parameters~\cite[Chapter~3]{xuArchitectureBlockchainApplications2019} and can be  optimised for a specific use-case~\cite[Chapter~6.3]{xuArchitectureBlockchainApplications2019}.
We envision future work to go beyond reporting gas cost and contribute to a discussion on the assumptions, advantages and drawbacks of different deployment options. Here, the question remains what properties must be investigated for a specific use case and which can be simulated or deduced from existing benchmarks, e.g.\ of the underlying blockchain platform.

\subsubsection{Enactment on public blockchains.}
From our cost analysis (Section~\ref{sec:methods}), we see that current transaction fees render the public Ethereum mainnet prohibitively expensive for the presented approaches in our primary studies. When considering blockchain, less quantifiable requirements play an important role also. Many of the guarantees a blockchain offers are a result of decentralisation~\cite[Chapter~3.2]{xuArchitectureBlockchainApplications2019}. Certain high-risk use cases (e.g., the transportation of dangerous goods, as in Meroni et al.~\cite{meroniTrustedArtifactDrivenProcess2019}) may benefit from decentralised and resilient public blockchains. Beyond Ethereum, we see a lot of promise in exploring alternative public blockchains. Next generation proof-of-stake blockchains like Algorand\footnote{\url{https://www.algorand.com}, accessed 2022-05-30.} or Avalanche\footnote{\url{https://www.avax.network}, accessed 2022-05-30.} promise a more sustainable operation and low cost. Studying enactment approaches on different public platforms would produce valuable insights. However, identifying advantages and drawbacks, and comparing different blockchain setups could prove to be challenging. We envision a first step in connecting our presented taxonomy to a taxonomy of blockchain platforms.

Beyond exploring alternative public blockchains, a different line of research has opened around \textit{Layer-2 technologies}.\footnote{Layer 2 scaling, Ethereum development documentation, \url{https://ethereum.org/en/developers/docs/scaling/##layer-2-scaling}, accessed 2022-05-30.}
These technologies reduce the involvement of the blockchain and perform most tasks off-chain; these off-chain tasks remain verifiable on the blockchain. We expect monitoring approaches to benefit in the short term from this line of research, as storing process traces becomes significantly cheaper. Long-term, we believe that enforcement approaches can make use of verifiable off-chain computations to significantly reduce cost. 
%
\section{Conclusion and Outlook}
We performed a SLR on blockchain-based business process enactment. We identified a final set of 36 primary studies. Based on these primary studies, we developed a taxonomy capable of describing the guarantees and capabilities of enactment approaches, and classified each study accordingly. 
We discussed our results in relation to the challenges of interorganisational processes. 

We find that blockchain is a natural fit to ensure traceability and enforce correctness of process execution. 
However, in terms of research focusing on flexibility and scalability, the field of blockchain-based enactment remains nascent. 
For all challenges, we have pointed out a range of research opportunities. We have not addressed privacy or security concerns as these are often a result of, or strongly dependent on, the employed blockchain technology. 
However, our taxonomy remains open for extensions---e.g., towards security or privacy properties. 
In the future, we envision the development of a decision model, based on our taxonomy, to support stakeholders considering blockchain-based enactment. 
%
%
\bibliography{bib}
%
\end{document}